\documentclass[twocolumn]{aa}
\usepackage{natbib}
\usepackage{color}
\usepackage{ragged2e}
\usepackage[pdfpagelabels=false]{hyperref}	% Hyperlinks
\hypersetup{colorlinks=true,linkcolor=blue,citecolor=blue,filecolor=blue,urlcolor=blue,}
\usepackage[varg]{txfonts}
\usepackage{graphicx,rotating}
\usepackage[normalem]{ulem}
\usepackage{colortbl}
\usepackage{xcolor}
\usepackage{bbold}

\usepackage{booktabs}
\usepackage{float}
\usepackage{bm}
\usepackage{caption}
\bibpunct{(}{)}{;}{a}{}{,} % to follow the A&A style

\definecolor{darkolivegreen}{rgb}{0.33, 0.42, 0.18}
\definecolor{salmon}{rgb}{0.95,0.5,0.25}

\begin{document}

\title{Constraints on the fuzzy dark matter mass using globular clusters in dwarf galaxies from Euclid ERO data}

\titlerunning{Constraints on the fuzzy dark matter mass using globular clusters in dwarf galaxies from Euclid ERO data}
\author{Lisa Besnard \inst{1}, Pierre Boldrini \inst{1}, Paola Di Matteo\inst{1}, Chervin Laporte\inst{1,3,4} and Teymoor Saifollahi\inst{2}}

\offprints{Pierre Boldrini,\\ \email{Pierre.Boldrini@obspm.fr}}
\institute{$^{1}$ LIRA, Observatoire de Paris, Université PSL, Sorbonne Université, Université Paris Cité, CY Cergy Paris Université, CNRS, 92190 Meudon, France, \\$^{2}$ Universit\'e de Strasbourg, CNRS, Observatoire astronomique de Strasbourg, UMR 7550, 67000 Strasbourg, France, \\
$^{3}$ Kavli IPMU (WPI), UTIAS, The University of Tokyo, Kashiwa, Chiba 277-8583, Japan, \\
$^{4}$ Institut de Ciències del Cosmos (ICCUB), Universitat de Barcelona, Martí i Franquès 1, E-08028 Barcelona, Spain}
\authorrunning{Besnard et al.}
\date{submitted to A$\&$A}

\abstract{We constrain the particle mass of the fuzzy dark matter model using globular cluster candidates in dwarf galaxies, from the Euclid Early Release Observations, within the Fornax and Perseus galaxy clusters. We model the orbital evolution of globular clusters by combining fuzzy dark matter density profiles with the corresponding dynamical friction formalism, and integrate their orbits using \texttt{galpy}. In the fuzzy dark matter framework, both the halo density profile and the dynamical friction force are modified by the wave-like nature of the ultra-light dark matter particle, whose mass is parameterized as $m_{22}=m_\chi/10^{-22},\mathrm{eV}$ and explored over the range $0.1 \leq m_{22} \leq 100$. This framework naturally alleviates the long-standing globular cluster timing problem originally identified in the Fornax dSph. We find that globular clusters no longer experience orbital decay toward the centers of their host galaxies for $m_{22}\leq1.46^{+0.51}_{-0.32}$ in the Fornax cluster and $m_{22}\leq1.09^{+0.14}_{-0.21}$ in the Perseus cluster. Future Euclid data releases, providing substantially larger samples of globular cluster candidates, will enable significantly tighter constraints on the dark matter particle mass within the fuzzy dark matter scenario.}

\maketitle

%%%%%%%%%%%%%%%%%%%%%%%%%%%%%%%%%%%%%%%%%%%%%%%%%%

%%%%%%%%%%%%%%%%% BODY OF PAPER %%%%%%%%%%%%%%%%%%

%GC, DF, D%

\section{Introduction}

A globular cluster (GC) is a compact, gravitationally bound group of stars. There are around 200 GCs in the Milky Way \citep{Gaia16,Garro24}, each one containing on average $10^5$-$10^6$ stars. These systems are spherical, very dense, and are situated everywhere in the Galaxy. The GC external dynamics raise an important issue in modern astrophysics : the timing problem of GCs in dwarf galaxies. Indeed, the cold dark matter (CDM) model, which has successfully described the behavior of dark matter (DM) on large scales for more than 20 years, faces several problems at galactic scales \citep{Bullock17,Boldrini21}, including the timing problem whose origin lies in dynamical friction (DF). First formulated in \cite{Chandra43}, DF slows down objects in orbits (such as GCs, star clusters, satellite galaxies) due to local gravitational interactions. The accumulation of smaller objects and matter behind the moving object leads to a gravitational force opposing the initial motion, making it lose energy and decelerate, eventually resulting in its infall toward the galactic center. This effect is amplified by the cuspy DM profile in galaxies predicted by CDM \citep{NFW96,NFW}. As a result, CDM simulations using the projected distance of the GCs from their host galaxy predict the migration of GCs into the galactic centers within a few Gyr. Yet, this prediction is in conflict with the observations as most GCs were formed more than 10 Gyr ago but are still present in galactic halos. This conflict between theory and observation is named the timing problem, and is particularly significant in dwarf galaxies, such as the Fornax dwarf spheroidal (dSph) \citep{Oh2000}, a satellite of the Milky Way, containing 5-6 GCs older than 10 Gyr \citep{Mackey03,Boer16,Wang19}. While dwarf galaxies are dominated by DM \citep{behroozi10,behroozi13}, we assume that their GCs contain no DM \citep{Ibata13} and thus are a useful probe of the dwarf galaxy’s DM density profile and can be used to investigate the timing problem. One of the most extensively explored solutions to the Fornax dSph timing problem is to modify the internal structure of the DM halo, determined by the cosmological model used, in order to reduce the efficiency of DF. \citep{Goerdt06,Read06,Cole12,Sanchez06,Inoue09,2018ApJ...868..134K,Meadows20,Boldrini19}.

A proposed solution to alleviate the GC timing problem is to consider alternative DM models, such as fuzzy dark matter (FDM) \citep{Goodman2000,Hu2000}. This ultra-light boson based model considers that DM is composed of a particle whose mass $m_{22}=m_\chi / 10^{-22}$ eV lies between 0.1 and 100, with $m_\chi$ the particle mass. Such a low mass induces quantum effects on galactic scales. The wave nature of FDM induces (i) dynamical heating leading to a reduction of DF \citep{Hui17,BO19,EZ20,Dutta21} and (ii) a modification of the DM density profile in dwarf galaxies \citep{Schive14}. These changes impact the orbits of the GCs, depending on the mass of the particle $m_{22}$, which is the only free parameter of the FDM model. At low particle masses ($0.1\lesssim m_{22}\lesssim3$), FDM effects are strongly pronounced, preventing the GCs from falling into the center of their host galaxy \citep{Hui17,BO19,Lancaster20}, whereas at higher masses ($30\lesssim m_{22}\lesssim100$), the model approaches the CDM behavior \citep{Hu2000}. Since the emergence of the timing problem, studies have focused on the Fornax dSph, and thus were limited by the small amount of observational data on dwarf galaxies hosting GCs. However, the Euclid mission has provided a large amount of observational data : almost 7000 GCs distributed among more than 800 dwarf galaxies in the local Universe \citep{saifollahi2025_perseus,saifollahi2025_fornax}. This includes the Fornax and Perseus galaxy clusters.  

A recent analysis based on Euclid Early Release Observations (ERO) \citep{ERO} investigated the GC timing problem within the standard CDM framework (\cite{Pierre2026}, in preparation). In this work, we extend this analysis to the FDM scenario in order to constrain the value of the FDM parameter $m_{22}$. The paper is structured as follows: Section 2 presents the theoretical framework and numerical methodology. We describe how the ultra-light nature of the FDM particle modifies both the DM density profile in dwarf galaxies and DF, and how we simulate the orbital evolution of the Euclid GC candidates using \texttt{galpy}. In Section 3, we first take the example of the Fornax dSph to illustrate FDM effects on GC dynamics. Then we use Euclid ERO data to compute the mass parameter $m_{22}$ for each dwarf galaxy in the Perseus and Fornax galaxy clusters by simulating the orbital evolution of their GCs, allowing us to provide an upper constraint on the FDM parameter $m_{22}$ above which FDM effects are no longer sufficient to prevent the GCs from falling into the center of their host galaxy. Finally, Section 4 summarizes our main results, discusses them in the context of previous studies, and presents perspectives for future work.

\section{Numerical method}
\label{fram}

In this section, we describe how the ultra-light nature of the FDM particle modifies both the DM density profile in dwarf galaxies and DF, and how we simulate the dynamics of Euclid GC candidates with the orbital integration methods, focusing on the role of the only free parameter of the FDM model $m_{22}$.

\subsection{Euclid globular cluster candidates}
\label{EuclidGC}

The GC candidate samples are drawn from the two Euclid ERO fields, the Fornax and Perseus clusters, presented by \citet{saifollahi2025_fornax} and \citet{saifollahi2025_perseus}, respectively. These samples include GC candidates associated with dwarf galaxies in the Fornax cluster (at about 20\,Mpc) in the Perseus cluster (at about 72\,Mpc). The initial ERO sample consists of 47 and 6653 GCs, distributed across 16 and 826 dwarf galaxies in the Fornax and Perseus clusters, respectively. We retain only clusters with $M_{I_{\rm E}} < -8$, imposing a lower mass limit and selecting only the more massive GCs. The GC masses are estimated using:
\begin{equation}
    M_{\mathrm{GC}} = 1.8 \times 10^{-0.4\,(M_{I_{\rm E}} + 0.5 - 4.82)} M_{\odot}.
    \label{Eq1}
\end{equation}
This equation is given an average $M/{L_V} = 1.8$ for old metal-poor single stellar populations \citep{2003MNRAS.344.1000B}, an average offset of -0.5 between V-band and $I_{\rm E}$-band for such stellar population (\citealp{saifollahi2025_fornax} in the appendix), and the V-band absolute magnitude of the Sun \citep{2018ApJS..236...47W}. This results in cluster masses in the range $1.5 \times 10^{5}$ to $3.1 \times 10^{6}\,M_{\odot}$ for the selected sample. In addition, we exclude objects with projected radii smaller than $0.2\,\mathrm{kpc}$, as these are more likely to be nuclear star clusters. Applying these criteria reduces our sample to approximately 4500 clusters across both environments, which still provides a statistically significant data set.

The positions of individual GCs are measured with sub-pixel accuracy, corresponding to uncertainties of only a few parsecs. Finally, we analyze a sample of 39 GC candidates in 12 dwarf galaxies in the Fornax galaxy cluster and 4487 GC candidates in 749 dwarf galaxies in the Perseus galaxy cluster. These Euclid GC candidates represent the surviving GC population observed today, excluding clusters that may have already undergone orbital decay to the galactic center.

\subsection{Fuzzy dark matter halo potential of dwarfs}
\label{FDMDP}
\begin{figure}[!t]
    \centering
    \includegraphics[width=1\linewidth]{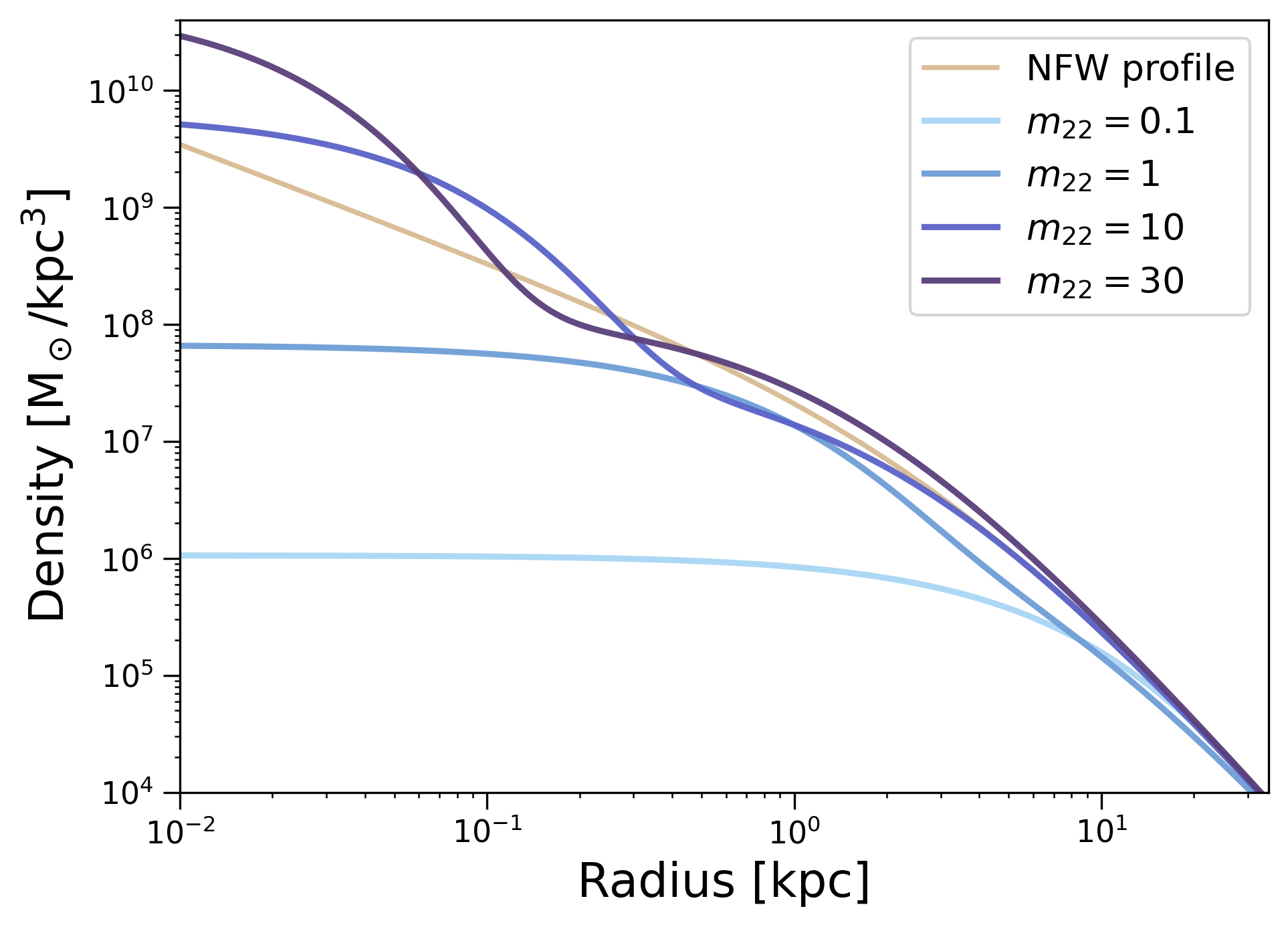}
    \caption{Comparison between CDM and FDM density profiles, assuming a NFW halo and a FDM  $10^{10} \;\mathrm{M}_\odot$ halo depending on the mass parameter $m_{22}$ as defined in Equations \ref{rhoc} and \ref{rhoe}.}
    \label{rho}
\end{figure}

To integrate the orbits of the GC candidates, we need to model the potential in which they evolve. The DM density profiles of the host dwarf galaxies depend on the cosmological model we use. In this section, we describe the DM density profile predicted by FDM. The ultra-light nature of the FDM particle induces a cored density profile, in contrast to the cusp profiles ($\rho \propto r^{-1}$) predicted by CDM \citep{NFW96,NFW}. A cored profile has an approximately constant central density ($\rho \approx \mathrm{const}$) and converges toward the CDM case in the outer region. We represent the CDM and FDM density profiles for different masses $m_{22}$ in Figure \ref{rho}.

The DM halo mass of the host galaxy is estimated from its stellar mass (provided by Euclid) using the stellar-to-halo mass relation of \cite{behroozi10,behroozi13} implemented in the \texttt{Python} \texttt{Halotools} package \citep{Hearin17}. We note, however, that this relation is poorly constrained and exhibits substantial scatter at the low halo masses relevant here, and that its use therefore involves a significant extrapolation for the least massive dwarfs in our sample. We neglect the stellar mass of the host galaxy, as it is negligible compared to the DM halo mass in dwarf galaxies \citep{behroozi10,behroozi13}. With this approximation, we underestimate the density of the galaxies and therefore of the DF force applied to the GCs. As a result, the upper constraint we obtain (see Section \ref{results}) on the FDM particle mass $m_{22}$ should be interpreted as an overestimated limit, since a more realistic case including the stellar mass would require a lower $m_{22}$ to compensate for the resulting stronger DF.

To model the density profile in dwarf galaxies in CDM, we assume a NFW halo defined in \cite{NFW}. The halo concentration parameter c is computed using the Prada concentration-mass relation \citep{Prada12} at redshift $z=0$, implemented in the \texttt{Python} \texttt{Colossus} package \citep{Diemer18}, from which we determine the scale radius $ r_{\rm s}$ \citep{Dutton14}. To model the FDM case, we construct the density profile as the sum of a central solitonic core and an outer power-law envelope (see Equations \ref{rhoc} and \ref{rhoe}), both implemented using the \texttt{TwoPowerSphericalPotential} class in \texttt{galpy}, as in \cite{Adrian26} and \cite{Pierre26}. 
\begin{equation}
\rho_{\rm core}(r,m_{22})=\frac{\rho_{01}(m_{22})} {4\pi r_{c1}^{3}(m_{22})}
\left(1+\frac{r}{r_{c1}(m_{22})}\right)^{-\beta}
\label{rhoc}
\end{equation}
\begin{equation}
\rho_{\rm ext}(r)=\frac{\rho_{02}}{4\pi r_{c2}^{3}}
\left(1+\frac{r}{r_{c2}}\right)^{-3}
\label{rhoe}
\end{equation}
$r_{c1}$ and $r_{c2}$ are respectively the characteristic radii of the core and outer components \citep{Schive14} while $\rho_{01}$ and $\rho_{02}$ represent their corresponding central densities. Finally, we use $\rho_{\rm FDM}= \rho_{\rm core}+\rho_{\rm ext}$. As the FDM model converges to the CDM model at large scales, this construction ensures a transition between the flat core and the CDM-like outer halo. We observe in Figure \ref{rho} the convergence of FDM with CDM in the outer region, and the expected FDM flat core for $m_{22}=0.1$ and $1$ in the inner region. This FDM profile modifies the orbit of the GCs compared to CDM. Indeed, at low masses ($m_{22}=0.1,1$), the density is lower compared to CDM because of the flat core, leading to a reduced DF force thus preventing the GCs from falling into their host galaxy. However, we find that for higher values of $m_{22}$ ($m_{22}=10,30$), the inner density of FDM is higher than for the CDM model, which may accelerate the fall of the GCs in this region. In summary, lower values of $m_{22}$ result in a larger characteristic core radius $r_c$, corresponding to a more uniform redistribution of DM within the galaxy. The resulting cored profile leads to reduced DF, and therefore to a longer infall time for an orbiting object.

\subsection{Dynamical friction in fuzzy dark matter}
\label{FDMDF}
In the FDM model, the ultra-light mass of DM particles induces a different DF behavior. Indeed, the wave nature of DM induces constructive and destructive interferences leading to density fluctuations (granules). These granules produce dynamical heating which reduces the efficiency of the DF force, thus increasing the infall time of the GCs. We use the quantum Mach number $\mathcal{M}_Q$ \citep{Lancaster20} to distinguish two regimes: 
\begin{equation}
    \mathcal{M}_Q = 44.56 \left( \frac{v}{1\, \mathrm{km\, s}^{-1}} \right) 
\left( \frac{M_{\mathrm{GC}}}{10^5 \mathrm{M}_\odot} \right)^{-1} m_{22}^{-1}
\end{equation}
where $v$ is the velocity of the GC and $M_{\mathrm{GC}}$ its mass. When $\mathcal{M}_Q \ll 1$, we use the classical DF whereas when $\mathcal{M}_Q \gg 1$, the FDM DF force is approximated \citep{Hui17,Lancaster20,BO19} by :
\begin{equation}
    \bm F_{\rm FDM} = -4\pi G^2 M_{\rm GC}^2 \rho(r)\,\frac{\bm{v}}{v^3}\,C_{\rm FDM}(kr,M_\sigma)
\end{equation}
where r is the galactrocentric distance of the GC and $\rho(r)$ the local density of the host galaxy. The coefficient $C_{\rm FDM}$ depends on two dimensionless parameters : (i) $kr$, with the wave number $k=\frac{m_\chi v}{\hbar}$ and (ii) the classical Mach number $M_\sigma=\frac{v}{\sigma}$, with $\sigma$ the local velocity dispersion of the host galaxy. Only the coefficient $C_{\rm FDM}(kr,M_\sigma)$ differs from the following classical expression : 
\begin{equation}
   \bm F_{\mathrm{CDM}}=-4\pi G^2 M_{\mathrm{GC}}^2 \rho(r) \frac{\bm{v}}{v^3} C_{\mathrm{CDM}}(r,v)\,
\end{equation}
With the $C_{\rm CDM}$ coefficient :
\begin{equation}
C_{\rm CDM}(r,v)=\frac{1}{2}\ln\!\left(1+\Lambda^{2}\right)\left[
\operatorname{erf}\!\left(\frac{v}{\sqrt{2}\sigma}\right)-\frac{2v}{\sqrt{2\pi}\sigma}
e^{-\frac{v^{2}}{2\sigma^{2}}}\right].
\end{equation}
We distinguish two sub-regimes of FDM DF : when  $M_\sigma \gg kr$, FDM halos have negligible velocity dispersion (static backgrounds), and the $C_{\rm FDM}$ coefficient is approximated \citep{Hui17,Lancaster20} by :

\begin{equation}
C_{\rm FDM}(kr)=\operatorname{Cin}(2kr)+\frac{\sin(2kr)}{2kr}-1
\label{CFDM1}
\end{equation}
with the cosine integral function :
\begin{equation}
\mathrm{Cin}(z) = \int_{0}^{z} \frac{1 - \cos(t)}{t}\, dt
\end{equation}
whereas when $M_\sigma \ll kr$, FDM halos have non-negligible velocity dispersion. In this case the $C_{\rm FDM}$ coefficient is approximated \citep{BO19,Lancaster20} by :
\begin{equation}
    C_{\rm FDM}(kr,M_\sigma)=\ln\!\left(\frac{2kr}{M_\sigma}\right)\left[\operatorname{erf}\!\left(\frac{v}{\sqrt{2}\sigma}\right)-\frac{2v}{\sqrt{2\pi}\sigma}e^{-\frac{v^{2}}{2\sigma^{2}}}\right].
\end{equation}

In summary, when $30 \lesssim m_{22}$, FDM DF approaches the classical regime, whereas when $m_{22} \lesssim 3$, the ultra-light nature of the particle strongly reduces DF \citep{Adrian26}. Therefore, we want to determine the maximum value of $m_{22}$ for which FDM effects remain sufficient to prevent the GC timing problem in dwarf galaxies.

\subsection{Orbital integration of globular clusters}
\label{galpy}
The \texttt{galpy} library is a public code \citep{Bovy15} developed for orbit integration in both \texttt{Python} and \texttt{C} and is widely used in galactic dynamics. \texttt{galpy} allows us to (i) model the CDM and FDM dwarf galaxies density profiles, assuming a spherical potential in both models, (ii) model FDM and CDM DF, and (iii) integrate the orbits of the GCs using the fast C integrator \texttt{dop853\_c}. We use the CDM DF using the classical formalism of \cite{Chandra43}, while the FDM DF is modeled using the \texttt{FDMDynamicalFrictionForce} class implemented in \texttt{galpy}, as introduced in \cite{Adrian26}. The orbits of the GCs are initialized at their observed present-day projected orbital radius $R_{\rm 2D}$, provided by the Euclid ERO data, and assuming a circular velocity which depends on the potential of the host galaxy. We will subsequently introduce a deprojected radius $R_{\rm 3D}$ and an eccentricity e for the GC orbits. A constant half-mass radius $r_{\rm hm}=10$ pc is assumed for all GCs. 

We adopt a reference timescale of 10 Gyr, approximately corresponding to mean ages of GCs. With the limit radius $r_{\rm lim}=0.2$ kpc, which separates the inner and outer regions of dwarf galaxies, these two parameters serve as a reference to determine whether a GC presents a timing problem. If the apocentric radius of a GC reaches $r_{\rm lim}=0.2$, we consider it to have fallen into its host galaxy, and if it falls within 10 Gyr, the GC is subject to the timing problem. This criterion allows us to compute the maximum FDM mass $m_{22}$, between 0.1 and 100, for which FDM effects are sufficient to prevent the GCs from falling into their host galaxy, thus for which the GCs are no longer subject to the timing problem. To do so, we integrate the orbit of a GC for several $m_{22}$, recomputing the corresponding DF and the DM density profile. To efficiently determine $m_{22}$, we use a bisection search. We start with $m_{22}=50$, the midpoint of our explored range. If the GC reaches 0.2 kpc within 10 Gyr, we test $m_{22}=25$, a lower value, to increase FDM effects which will prevent the GC from falling. Otherwise we test $m_{22}=75$. Repeating this algorithm iteratively allows us to converge on the maximum value of $m_{22}$ for which the GC does not fall into the center of its host galaxy. The bisection method assumes a monotonic dependence of the orbital evolution of the GCs on $m_{22}$. The corresponding computational time is approximately 48 CPU hours for the Perseus galaxy cluster ($\sim $ 2000 GCs).

\begin{figure}[!t]
    \centering
    \includegraphics[width=1\linewidth]{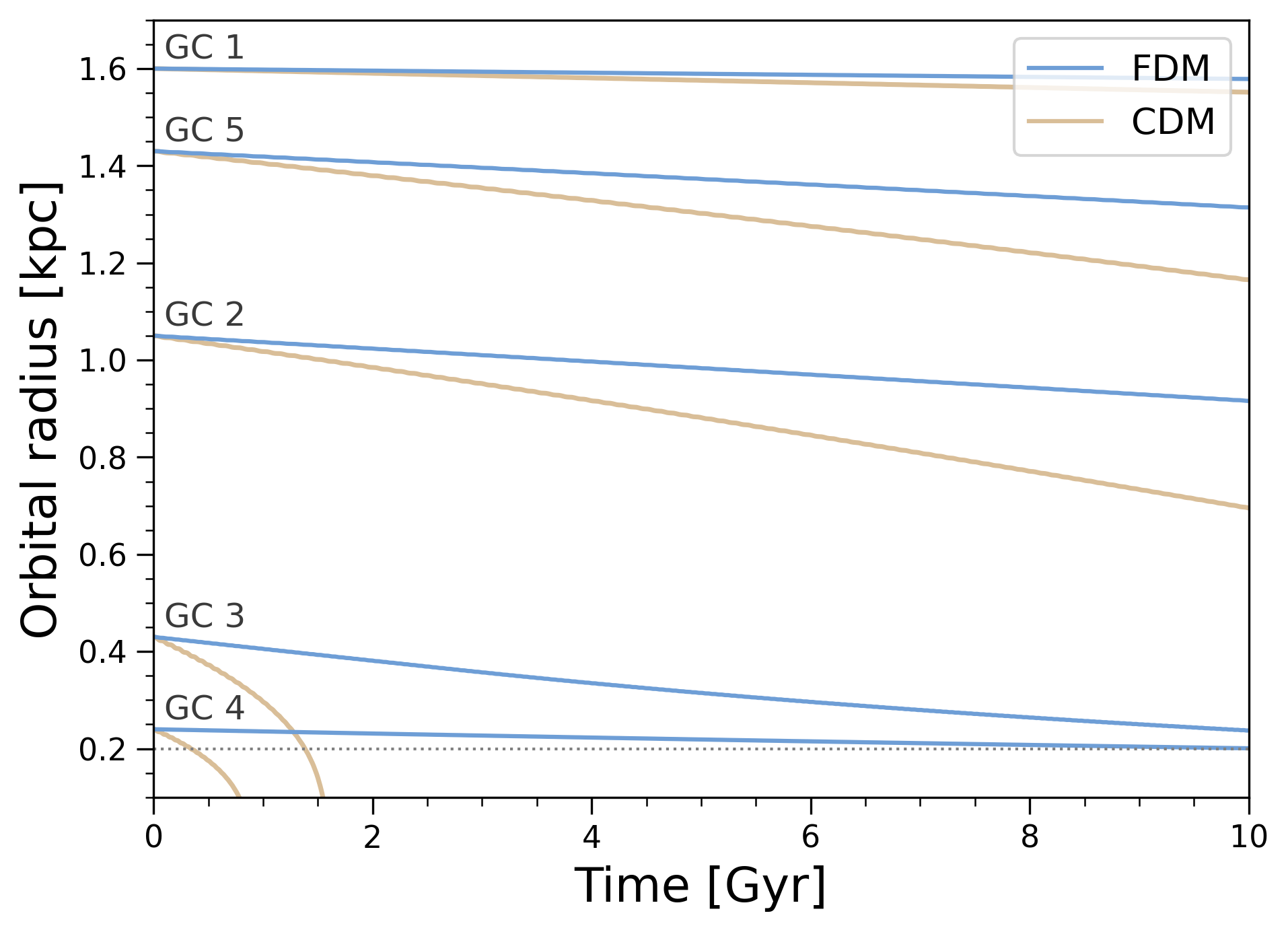}
    \caption{Orbital radius as a function of time of the five Fornax GCs. Beige lines show the GCs trajectories in the CDM model while blue lines correspond to the FDM model, with a modified DF, a cored DM density profile and $m_{22}=1.33$ for which none of the five Fornax GCs falls below 0.2 kpc within 10 Gyr. We assume a stellar mass of $M_\star = 2.39 \times 10^7 \; \mathrm{M_{\odot}}$ and we use the initial projected radius $R_{\rm 2D}$.}
    \label{fornax}
\end{figure}

\section{Results}\label{results}

In this section we simulated the orbital evolution of GCs with \texttt{galpy}, including FDM DF and FDM density profile described in section \ref{fram}. We provide an upper constraint on the FDM parameter $m_{22}$ above which FDM effects are no longer sufficient to prevent the GCs from falling into the center of their host galaxy. We first illustrate the FDM effects on the Fornax dSph, and then we generalize to the local Universe using Euclid data.

\subsection{Fornax dwarf galaxy globular clusters}
\label{fornaxGC}

We use the example of the Fornax dSph and its five GCs, to illustrate the timing problem in CDM, and how FDM provides a solution to it depending on the mass parameter $m_{22}$. We integrate the orbits of the five Fornax dSph GCs given their present-day projected orbital radii $R_{2D}$ and their masses taken from \cite{Boer16}, taking into account FDM DF and the FDM cored density profile which both depend on $m_{22}$. We compare the Fornax dSph GCs trajectories in FDM and CDM over 10 Gyr in Figure \ref{fornax}. The DM halo mass of the Fornax dSph is estimated from its stellar mass $M_\star = 2.39 \times 10^7 \; \mathrm{M_{\odot}}$, taken from \cite{Huang21}. Figure \ref{fornax} shows that the CDM model predicts that GC3 and GC4 should fall in the Fornax dSph within less than 2 Gyr, which is inconsistent with their observed ages exceeding 10 Gyr, leading to a timing problem. Conversely, in both the FDM and CDM models, GC1, GC2 and GC5 do not fall into the galaxy center and are therefore not subject to the timing problem.

In the FDM model, both the core profile and DF prevent GC3 and GC4 in the Fornax dSph from reaching its center, for a maximum mass parameter of $m_{22}=1.33$. For $m_{22}>1.33$, the FDM DF and DM density profile are no longer sufficient to prevent GC4 from falling into Fornax. Furthermore, for $m_{22}>1.54$, GC3 also falls. On the contrary, for $m_{22} \leq 1.33$, FDM effects are strengthened and still prevent GC3 and GC4 from falling within 10 Gyr. This results in an upper constraint on $m_{22}$ for the Fornax dSph obtained from the most constraining GC (here GC4). Finally, for $m_{22} \leq 1.33$ none of the Fornax dSph's GCs have a timing problem.
\begin{figure}[!t]
    \centering
    \includegraphics[width=1\linewidth]{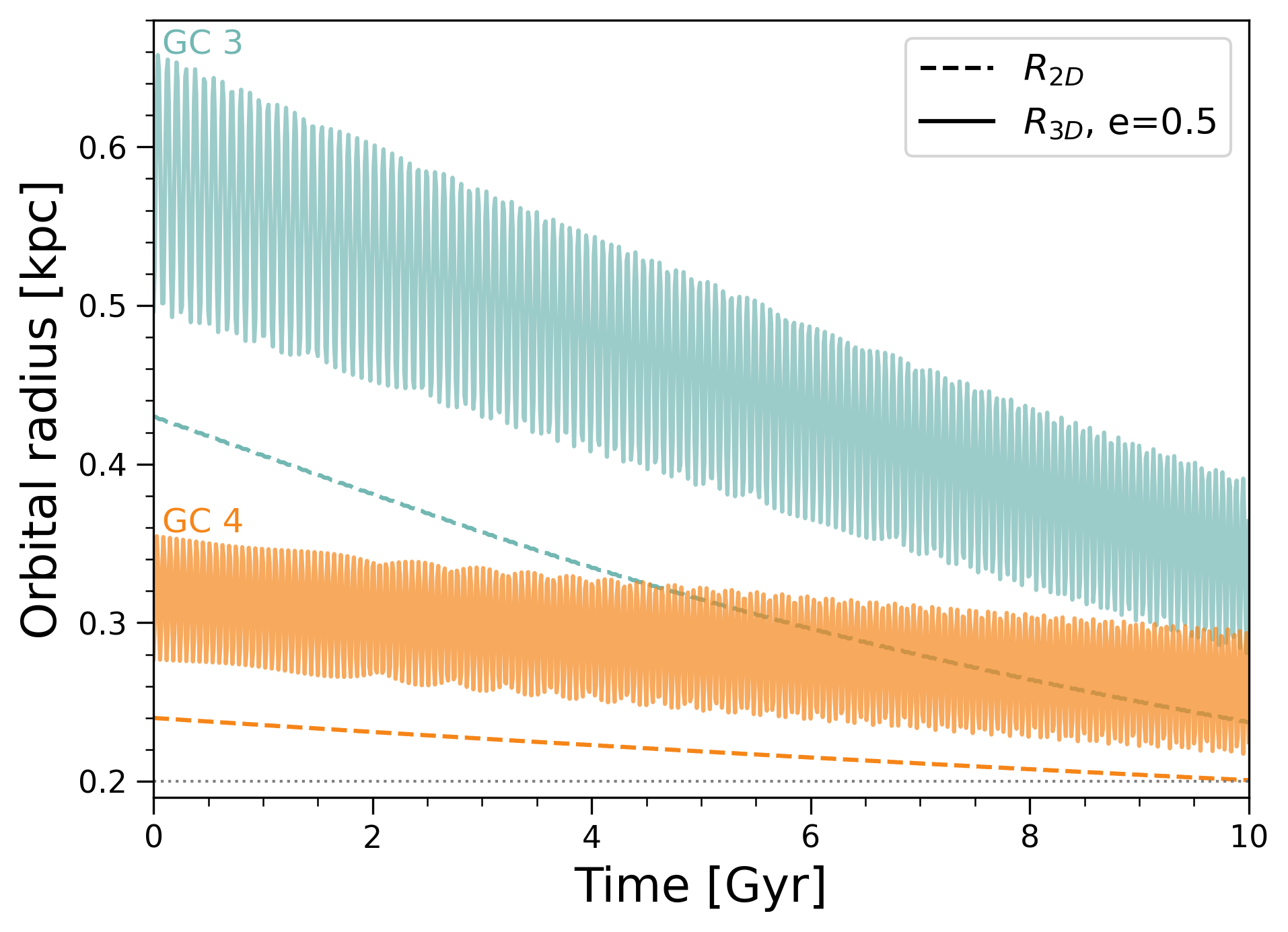}
    \caption{Orbital radius as a function of time for GC3 and GC4 of the Fornax dSph. $R_{\rm 3D}$ is the correction of the projected radius $R_{\rm 2D}$. This correction and the introduction of eccentricity $e$ both increase the infall time of the GCs. Here, $m_{22}=1.33$.}
    \label{GC34}
\end{figure}
To improve the modeling of the GCs trajectories, we consider a range of orbital configurations. First, we estimate a 3D radius $R_{\rm 3D}=\frac{2}{\sqrt{3}} R_{\rm 2D}$, since $R_{\rm 2D}$ is the projected distance of the GC orbital radius. A major uncertainty, however, is whether the present-day radii of the Euclid GCs are representative of their birth radii, which remain unconstrained. In the specific case of the Fornax dSph, previous studies have shown that assuming larger initial galactocentric radii can substantially increase the orbital decay timescales and alleviate, or even remove, the GC timing problem \citep{Cole12,Boldrini19,Meadows20}. Whether a similar scenario applies to the GC populations observed by Euclid remains unknown, as their birth-radius distribution is currently unconstrained. By adopting the deprojected present-day radius $R_{\rm 3D}$ as our reference initial radius, we maximizes the timing problem and therefore provides the lowest $m_{22}$ constraint within our framework. Secondly, we include an orbital eccentricity $e$. Finally, we combine these two parameters ($R_{\rm 2D}$ or $R_{\rm 3D}$ and $e$) to cover a wide range of configurations. We adopt $e=0$ and $e=0.5$, respectively corresponding to a circular orbit with the fastest orbital decay, and to a realistic eccentric orbit used to assess the impact of orbital eccentricity. Figure \ref{GC34} shows the impact of adding eccentricity and correcting the projected radius on GC3 and GC4 of the Fornax dSph. With the radius $R_{\rm 3D} > R_{\rm 2D}$, the GCs are located farther from the galaxy center, in a lower-density region leading to weaker DF. Adding eccentricity increases the GCs's velocity, slowing down their fall into the host galaxy. A GC with eccentricity $e =0.5$ is considered to have reached the center of its host galaxy when its last apocenter (in a timescale of 10 Gyr) is below 0.2 kpc. In summary, a realistic configuration of the orbits ($R_{\rm 3D}$ and $e=0.5$) increases the infall time of the GCs, leading to a higher particle mass needed to prevent their fall ($m_{22}\leq2.24$) compared to $R_{\rm 2D}$ and $e=0$ ($m_{22}\leq1.33$).

\subsection{Euclid globular clusters of dwarf galaxies in the Fornax and Perseus galaxy clusters}

In this section, we constrain the value of $m_{22}$ of the respectively 12 and 749 dwarf galaxies of the Fornax and Perseus galaxy clusters using Euclid data on their GCs. First, we identify the GCs with a timing problem in CDM, by integrating their orbits with \texttt{galpy} in a CDM potential using classical DF. Each GC is simulated four times, under the four configurations defined in section \ref{fornaxGC} : ($R_{\rm2D}$, e=0), ($R_{\rm2D}$, e=0.5), ($R_{\rm3D}$, e=0), ($R_{\rm3D}$, e=0.5). For each configuration, we select only GCs that reach 0.2 kpc from the center of their host galaxy within 10 Gyr. A GC with eccentricity is considered to have fallen into its host galaxy when its last apocenter is below 0.2 kpc after 10 Gyr. The number of GCs satisfying these criteria varies depending on the orbital configuration. For instance, we use a set of data of 9 GCs in 7 galaxies in the Fornax galaxy cluster and 1865 GCs in 288 galaxies in the Perseus galaxy cluster, composed of all the GCs that are subject to the timing problem in the CDM model, under the configuration $R_{\rm 2D}$, e=0.5. Then, we simulate the orbits of the selected GCs in an FDM potential, including FDM DF, to determine, for each orbital configuration, the value of $m_{22}$ above which the GC no longer sinks to the center of its host galaxy.

\begin{figure}[!t]
    \centering
    \includegraphics[width=1\linewidth]{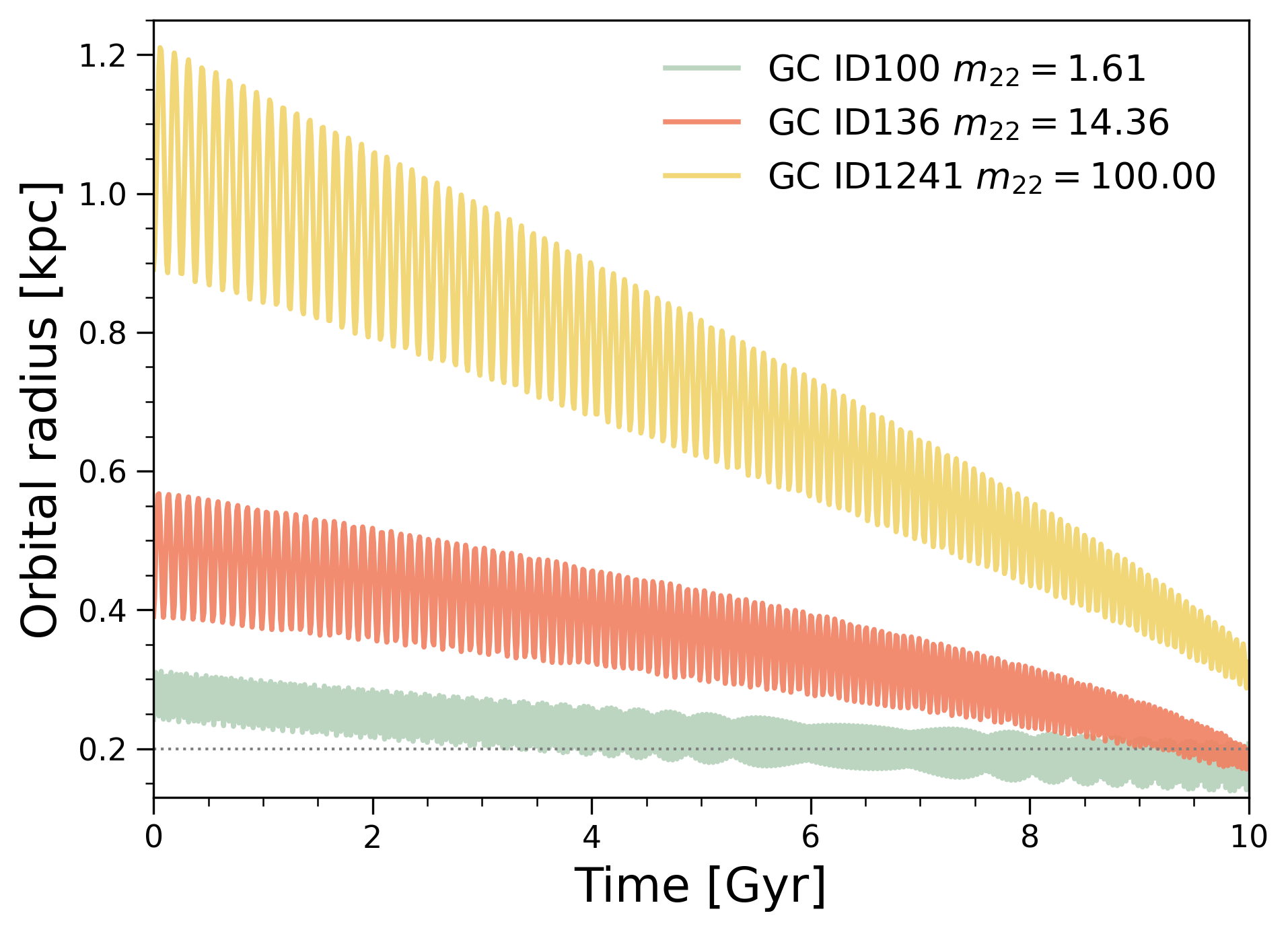}
    \caption{Orbital radius as a function of time for three GCs in dwarf galaxies in the Perseus galaxy cluster, integrated using the maximum $m_{22}$ for which none of them falls below 0.2 kpc within 10 Gyr. We use eccentricity $e=0.5$ and the corrected radius $R_{\rm3D}$. A GC with eccentricity $e =0.5$ is considered to have reached the center of its host galaxy when its last apocenter (in a timescale of 10 Gyr) is below 0.2 kpc}
    \label{3gal}
\end{figure}
The trajectories of three GCs in three different galaxies from the Perseus galaxy cluster are shown in Figure \ref{3gal}. We use the corrected radius $R_{\rm 3D}$, the eccentricity $e=0.5$ and the maximum values of $m_{22}$ for which the FDM density profile and FDM DF are sufficient to prevent their fall into the center of their host galaxy (0.2 kpc) within 10 Gyr. Figure \ref{3gal} shows that for different initial conditions depending on the orbital radius and the GC's mass, a GC requires a different value of $m_{22}$, which determines both the FDM galaxy potential and DF such that the GC does not fall into the center of its host galaxy. Some GCs (as shown for the yellow trajectory) require $m_{22}=100$, which is the upper limit of our explored range (0.1 to 100). In these cases, FDM effects are sufficient to maintain their orbits compared to the CDM case, regardless of the value of $m_{22}$ in the range 0.1–100. Thus, the maximum mass $m_{22}$ for which they do not fall into the center of their host galaxy is the upper bound of our calculations. 

\begin{figure}[!t]
    \centering
    \includegraphics[width=1\linewidth]{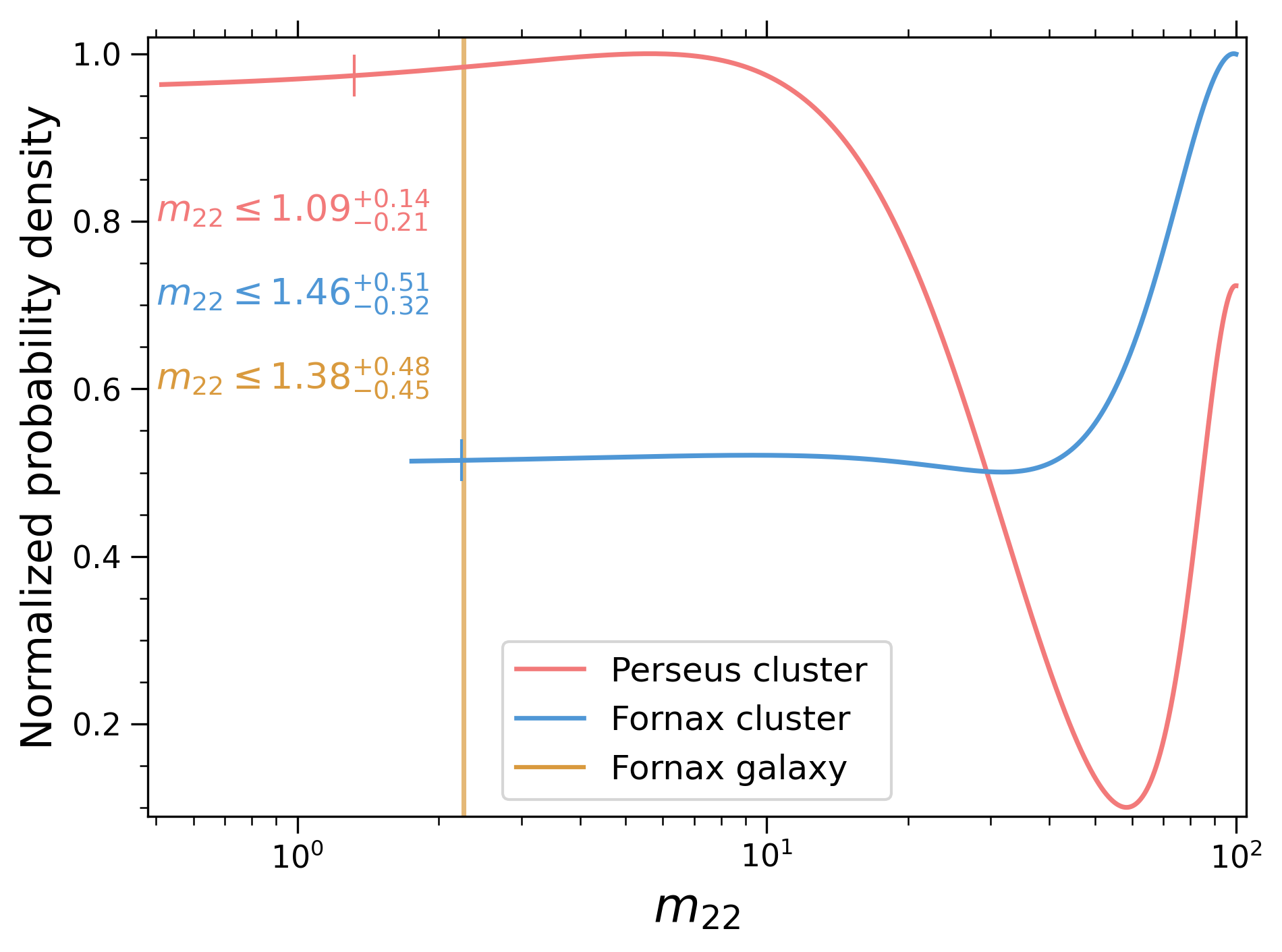}
    \caption{Normalized probability density of the FDM parameter $m_{22}$ for dwarfs of the two galaxy clusters, and the value for the Fornax dSph, using the corrected radius $R_{\rm3D}$ and the eccentricity $e=0.5$, where the tick marks indicate the corresponding $5^{th}$ percentile. The constraints on $m_{22}$ including all orbital configurations (where each GC is simulated with and without eccentricity, and with $R_{\rm2D}$ and $R_{\rm 3D}$) are written on the left.}
    \label{m22}
\end{figure}

We then determine, for each galaxy of the Fornax and Perseus galaxy clusters, the maximum value of $m_{22}$ (between 0.1 and 100) for which FDM effects are sufficient to ensure that none of the galaxy's GCs fall into the central region. This value corresponds to the most constraining GC in each galaxy, which has the minimum $m_{22}$. This results in a distribution of $m_{22}$ values, with each value corresponding to one dwarf galaxy. The value of $m_{22}$ computed for the Fornax dSph and the distributions of $m_{22}$ obtained for the Fornax and Perseus galaxy clusters are represented in Figure \ref{m22}. The values of $m_{22}$ are computed using only the corrected radius $R_{\rm 3D}$ and the eccentricity $e = 0.5$.

We constrain $m_{22}$, respectively for the Fornax and Perseus galaxy clusters, by computing the $5^{th}$ percentile of their corresponding distribution. We do not constrain $m_{22}$ with the minimum value of the distribution to avoid a final constraint driven by only a single GC. Thus, we give an upper constraint on $m_{22}$ determined by the lowest $5\%$ of the distribution, below which we consider that the FDM model provides a solution to the timing problem. In Figure \ref{m22}, the constraints of $m_{22}$ for the Fornax and Perseus galaxy clusters are shown as tick marks. They are based on the distributions computed using only the corrected radius $R_{\rm3D}$ and the eccentricity $e = 0.5$. To account for the impact of the different orbital configurations on the constraint (with or without eccentricity, and with $R_{2D}$ or $R_{3D}$, see Section \ref{fornaxGC}), we compute four separate distributions of $m_{22}$, each corresponding to one orbital configuration applied to the GCs : ($R_{\rm2D}$, e=0), ($R_{\rm2D}$, e=0.5), ($R_{\rm3D}$, e=0), ($R_{\rm3D}$, e=0.5). These four distributions are then combined, and we compute the $5^{th}$ percentile of the resulting distribution to obtain the final constraint on $m_{22}$, which accounts for the impact of the different orbital configurations. This calculation is performed for the Fornax and Perseus galaxy clusters and for the Fornax dSph respectively. The resulting constraints are reported on the left of Figure \ref{m22}. 
The uncertainty in the final constraint (induced by the four different orbital configurations) is estimated by computing the $5^{th}$ percentile of the four distributions separately (that each corresponds to one orbital configuration). The upper (lower) uncertainty is defined as the difference between the maximum (minimum) and the median of these four values. This calculation is, again, performed for the Fornax and Perseus galaxy clusters and for the Fornax dSph respectively. The resulting uncertainties are relatively small, highlighting that the different orbital configurations do not significantly alter the result, leading to similar constraints in $m_{22}$ and strengthening its reliability.

Finally, for $m_{22} \leq 1.46^{+0.51}_{-0.32}$ and $m_{22}\leq  1.09^{+0.14}_{-0.21}$, the GCs in the Fornax and Perseus galaxy clusters, respectively, are no longer subject to the timing problem, as the GCs of the Fornax dSph for $m_{22} \leq 1.38^{+0.48}_{-0.45}$. It provides three upper constraints in close agreement on the only free parameter of FDM.

Two studies based on the stellar dynamics of the Fornax and Sculptor dSphs have derived upper limits of $m_{22}<1$ \citep{Marsh15} and $m_{22}<0.4$ \citep{Gonzalez17}, respectively, which are broadly consistent with, although more restrictive than, our results. This difference is likely explained by the fact that these studies only account for the modification of the host galaxy density profile in the FDM framework, while neglecting the corresponding modification of DF. Other astrophysical probes constrain the FDM particle mass using independent methods. For example, \citet{Amorisco18} inferred $m_{22}>1.5$ from the survival of cold stellar streams, \citet{Chiang23} obtained $m_{22}\simeq0.5$--$0.7$ from the structure of the Milky Way disk, and several studies based on the Lyman-$\alpha$ forest derived significantly larger values, $m_{22}\simeq7$--$20$ \citep{Irsic17,Kobayashi17,Nori19,Armengaud17,Rogers21}. Our constraints are compatible with those inferred from the Milky Way disk, but are in tension with the lower bound from stellar streams and with the Lyman-$\alpha$ forest constraints.

\section{Conclusion}

In this work, we have constrained the value of $m_{22}$, the mass of the DM particle, under the FDM scenario, using GC candidates of dwarf galaxies identified in Euclid ERO data. $m_{22}$ is the only free parameter of the FDM model, an alternative model to CDM, composed of ultra-light bosons. The FDM model induces different galactic dynamics motivated by the small-scale challenges faced by CDM \citep{Bullock17,Boldrini21}. Indeed, CDM cosmological simulations predict that DF should cause the fall of the GCs into the center of their host galaxy within a few Gyr. However, this prediction is in conflict with the observations. This conflict is called the timing problem. We focused on resolving the timing problem for GCs in dwarf galaxies, taking into account both DF and cored density profiles induced by FDM. 

The Euclid ERO data provide access to an unprecedented sample of GCs ($\sim5000$) in two galactic cluster environments, allowing us to provide three constraints in close agreement for which the GCs no longer fall into the center of their host galaxy compared to CDM, assuming a FDM scenario : $m_{22} \leq 1.46^{+0.51}_{-0.32}$, $m_{22}\leq 1.09^{+0.14}_{-0.21}$ and $m_{22} \leq1.38^{+0.48}_{-0.45}$, for the Fornax and Perseus galaxy clusters and the Fornax dSph respectively. The errors are relatively small, strengthening the reliability of the result. Overall, our constraints are broadly consistent with those inferred from the Milky Way disk and stellar dynamics of nearby dwarf galaxies, while remaining in tension with constraints derived from stellar streams and the Lyman-$\alpha$ forest. 

In this work, we deliberately assume that the GCs evolved from their present-day galactocentric radii, thereby maximizing the timing problem. The resulting FDM constraint should therefore be interpreted as the lowest $m_{22}$ implied under this assumption. If the GCs instead formed at larger radii and subsequently migrated inward, the timing problem would be weakened, shifting the inferred constraint toward higher $m_{22}$ values and potentially reducing the tension with independent constraints on FDM. However, the initial radii cannot be increased arbitrarily. Such large inferred birth radii may be difficult to reconcile with the physical conditions required for massive cluster formation. GCs are expected to form preferentially in regions where the gas density and pressure are sufficiently high \citep{Elmegreen97}, while the dense environments required for their formation can also lead to rapid cluster disruption \citep{Kruijssen15}. GC formation and survival are therefore expected to favour a restricted range of galactocentric radii. Determining how far the FDM constraint can shift toward higher $m_{22}$ values ultimately requires independent constraints on the plausible birth-radius distribution of GCs in dwarf galaxies. Thus, the impact of different initial orbital conditions, particularly the GC birth-radius distribution, should be further explored to quantify their effect on the GC timing problem and, consequently, on the inferred constraints on $m_{22}$.

Future work should focus on using the large statistical samples provided by next Euclid data releases to extend this work to other dwarf galaxies and their GCs in new environments, and to assess the impact on the constraint on $m_{22}$. The numbers of robust GC candidates with high signal-to-noise photometry expected to lie within the Euclid observed regions out to 70 Mpc will approach $10^5$ \citep{Lancon21,Voggel25}. Another extension of this work could be to combine the FDM effects (DF, cored profiles) with a non-spherical potential for dwarf galaxies. In this work, we have assumed a spherical potential, although dwarf galaxies are expected to have non-spherical geometries in both CDM \citep{Hayashi_2012} and FDM \citep{Dome23}, which modifies the density profile of the host galaxy, the DF force and thus the dynamics of the orbiting GCs, leading to a possibly higher constraint on $m_{22}$.

Finally, a similar approach based on GC dynamics could also be applied to other DM models, such as warm dark matter (WDM) \citep{Bode01,Bond80,Dodelson94,Hogan00,Abazajian06,Viel05} or self-interacting dark matter (SIDM) \citep{Spergel00,Carlson92,deLaix95}, to constrain their parameter: the WDM particle mass $m_{\rm \nu}$ (in keV) and the SIDM cross-section $\sigma_m$ (in cm$^2/$g).

\section{Data Availability}

The data underlying this article is available through reasonable request to the author.

\begin{acknowledgements} 
PB and TS acknowledge funding from the CNES post-doctoral fellowship program. This work was supported by CNES, focused on the Euclid mission. PB is grateful to the Action Thématique de Cosmologie et Galaxies (ATCG), Programme National ASTRO of the INSU (Institut National des Sciences de l’Univers) for supporting this research, in the framework of the project "Coevolution of globular clusters and dwarf galaxies, in the context of hierarchical galaxy formation: from the Milky Way to the nearby Universe" (PI: G. Pagnini). 

\end{acknowledgements}

\bibliography{src}

\end{document}